\documentclass{PoS}

\usepackage{psfrag}
\usepackage{wrapfig}

\title{Exclusive diffractive Higgs and jet production\\at the LHC}

\ShortTitle{Exclusive diffractive Higgs and jet production at the LHC}

    \author{\speaker{Christophe Royon}\\%
    CEA/IRFU/Service de physique des particules, CEA/Saclay\\
    E-mail: \email{christophe.royon@cea.fr}}

    \author{\speaker{Rafal Staszewski}\\%
    CEA/IRFU/Service de physique des particules, CEA/Saclay\\
    Institute of Nuclear Physics, Polish Academy of Sciences, Krakow\\
    E-mail: \email{rafal.staszewski@ifj.edu.pl}}

    \author{Alice Dechambre\\
    IFPA, Dept. AGO, Universit\'e de Li\`ege\\
    E-mail: \email{alice.dechambre@ulg.ac.be}}

    \author{Oldrich Kepka\\
    Center for Particle Physics, Institute of Physics, Academy of Science, Prague\\
    E-mail: \email{kepkao@fzu.cz}}

\abstract{ The implementation of exclusive diffractive production of Higgs
boson and dijets in the FPMC (Forward Physics Monte Carlo) framework is
presented following the models by Khoze, Martin, Ryskin and Cudell, Dechambre,
Hernandez and Ivanov. The predictions of the models are compared to the CDF
measurement of exclusive jets and the uncertainties on the Higgs boson
and jet production cross sections at the LHC are discussed.  } 

\FullConference{XVIII International Workshop on Deep-Inelastic Scattering and Related Subjects\\
                April 19 -23, 2010\\
                Convitto della Calza, Firenze, Italy}

\begin{document}

\newcommand{\sidebyside}[5]
{
\begin{figure}[#1]
  \begin{minipage}[t]{0.47\linewidth}
    \centering
    \includegraphics{#2}
    \label{#2}
    \caption{#3}
  \end{minipage}\hfill%
  \begin{minipage}[t]{0.47\linewidth}
    \centering
    \includegraphics{#4}
    \label{#4}
    \caption{#5}
  \end{minipage}
\end{figure}
}

\section{Introduction} 

The Higgs boson is the last particle of the Standard Model remaining to be
discovered. Inclusive searches have been performed at the
Tevatron and are being started at the LHC.  However the search for the Higgs
boson at low mass is complicated -- depending on the decay channel -- either
due to the low branching ratio or due to the huge background coming from QCD
jet events. Thus other possibilities have been investigated, in particular
using the exclusive diffractive production. In such processes both incoming
hadrons, $p\bar p$ at the Tevatron and $pp$ at the LHC, remain intact after the
interaction and the Higgs decays in the central region. The process involves
the exchange of a color singlet, thus large rapidity gaps can remain between
the Higgs and the outgoing hadrons. Other particles, or systems of particles,
can also be produced, \textit{e.g.} a pair of jets. The great advantage of such
production mechanism is the possibility to detect fully exclusive events by tagging
both outgoing hadrons. This can lead to good mass resolution and bakground
rejection.

\section{Theoretical models}

The exclusive production can be modeled within QCD. In the simplest case the
process can be described as a two-gluon exchange -- one gluon involved in the
production and the other one screening the color (\textit{e.g} Fig. 1).  Such
calculation is well understood and under theoretical control, however to make
the description realistic following corrections need to be added: impact
factor, Sudakov form factor and rapidity gap survival probability.

The impact factor~\cite{Ivanov:2004ax} regulates the infra-red divergence and
embeds quarks inside the proton.  It is modeled phenomenologicaly and includes
soft physics. The Sudakov form factor~\cite{Khoze:2000mw} corresponds to virtual
vertex corrections and depends on two scales -- the hard scale linked to the
hard subprocess ($gg \to X$) and the soft scale related to the transverse
momentum of the active gluons -- the scale from which a virtual parton can be
emitted.  The Sudakov form factor suppresses the cross section by a factor of
the order of 100 to 1000. Finally, additional soft interactions of initial and
final state protons can occur~\cite{Frankfurt:2006jp}, which are taken into
account by introducing the rapidity gap survival probability.

In this work we study two models of exclusive Higgs and jets production: the
Khoze, Martin and Ryskin (KMR) model~\cite{Khoze:2000mw,Khoze:2001xm} and the
Cudell, Hern\'andez, Ivanov, Dechambre exclusive (CHIDe)
model~\cite{Cudell:2008gv}. The models are in fact very similar -- both use
perturbative calculations and have similar ingredients. However they differ in
details, which leads to different predictions.

\begin{wrapfigure}{r}{5cm}
  \centering
  \psfrag{p}{$p$}
  \psfrag{sgg}{$s_{gg}$}
  \psfrag{kT}{$k_T$}
  \includegraphics[width=4cm]{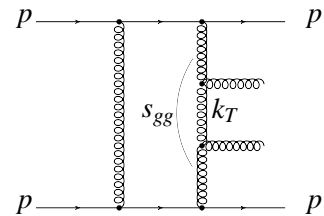}
  \label{diagram}
  \caption{Feynman diagram of exclusive jet production.}
  \vspace{-5mm}
\end{wrapfigure}

There are three main differences between the KMR and CHIDe models. The first
difference is the collinear approximation used in the KMR model contrary to the exact
kinematics used in CHIDe.  The second one is the variable used as the upper
scale of the Sudakov form factor in the exclusive jet case. It is chosen as the
gluon-gluon invariant mass, $s_{gg}$, in the KMR model, whereas in the CHIDe model
the transverse momentum squared of the gluon, $k_T$, is used (see Fig.~1).  The
last difference is the impact factor in CHIDe model that suppresses very soft
gluon emissions from the proton, which is not present in the KMR model.

\section{FPMC -- Forward Physics Monte Carlo}

The Higgs and jet exclusive production in both KMR and CHIDe models have been
implemented in the Forward Proton Monte Carlo (FPMC)~\cite{FPMC}, a generator
that has been designed to study forward physics, especially at the LHC. It aims
to provide the user a variety of diffractive processes in one common framework,
\textit{e.g.} the following processes have already been implemented: single
diffraction, double pomeron exchange, central exclusive production and
two-photon exchange.

The implementation of the KMR and CHIDe models in FPMC allows their direct
comparison using the same framework. In Fig.~2, we present the cross section of
exclusive Higgs boson production at the LHC as a function of the Higgs boson
mass.  In addition, we show the predictions from the KMR original
calculation~\cite{Khoze:2001xm} and the results of the implementation of the KMR
model in the ExHuME generator~\cite{Monk:2005ji}. The difference in results
between the FPMC and ExHuME implementations of the KMR model is the effect of
two factors. First, in ExHuME the value of the gluon distribution is frozen for
small $Q^2$, whereas in FPMC it vanishes to 0. The other reason of the
disagreement is a different implementation of the $gg \rightarrow H$ vertex --
in FPMC the HERWIG implementation is used whereas in ExHuME the vertex is
directly implemented. 

%  \sidebyside{h}
%  {models} {Cross section for exclusive Higgs production at the LHC as a function
%  of the Higgs mass. Results of CHIDe and KMR implemented in FPMC are presented.
%  For comparison the predictions of the original KMR model\cite{Khoze:2001xm}
%  and ExHuME generator are given.}
%  {CDF}{Exclusive jets production cross section at the Tevatron as a function of
%  the minimum jet $E_T$ threshold. The CDF measurements are compared to the CHIDe
%  and KMR models displayed after applying the jet algorithm.}

\sidebyside{h}
{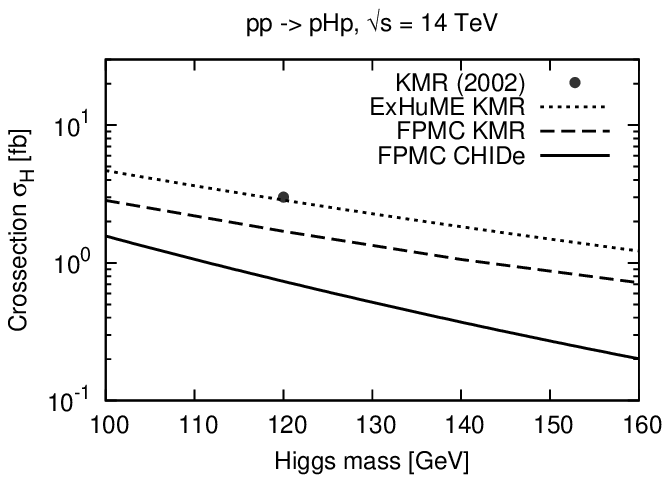} {Cross section for exclusive Higgs production at the LHC for various models.}
{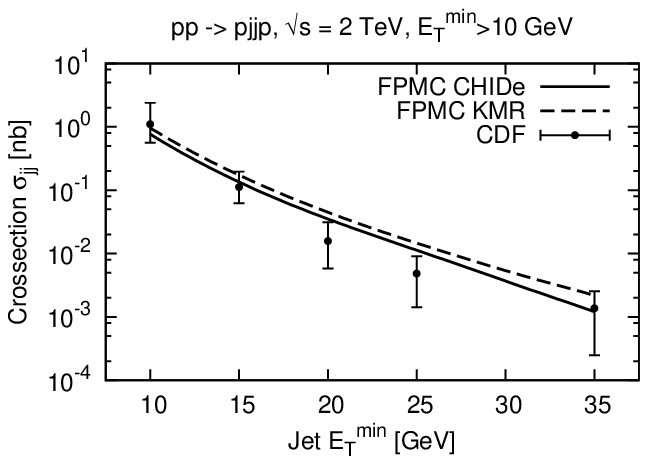}{Exclusive jets production cross section at the Tevatron as a function of jets $E_T^{min}$.}

The predictions of the KMR and CHIDe models are compared to the CDF measurement
of exclusive jets production at the Tevatron (Fig.~3). A good agreement
is found between the CDF measurement and the predictions of both CHIDe and KMR
models. The difference between the models is small compared to the data
uncertainties. 

\section{Uncertainties of the models}

% \sidebyside{b}
% {GluonsLhcHiggs}
% {Uncertainty of the CHIDe model due to the gluon distributions for exclusive
% Higgs production at the LHC.}
% {LowerScaleLhcHiggs}
% {Uncertainty of the CHIDe model due to the lower limit of integration in the
% Sudakov form factor for exclusive Higgs production at the LHC.}

\sidebyside{t}
{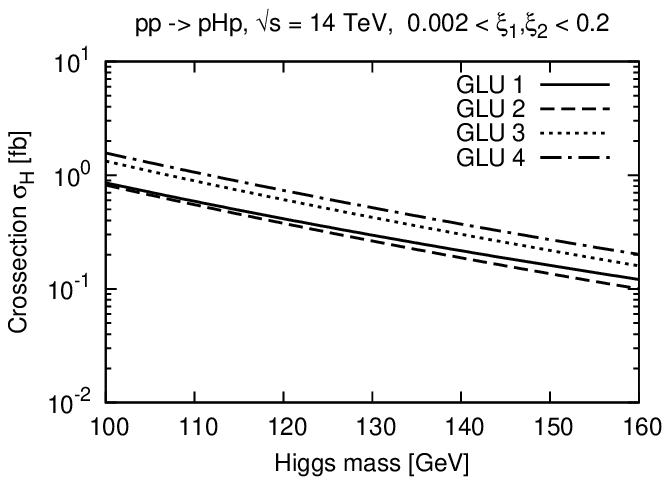}
{Uncertainty due to the gluon distributions for exclusive
Higgs at the LHC.}
{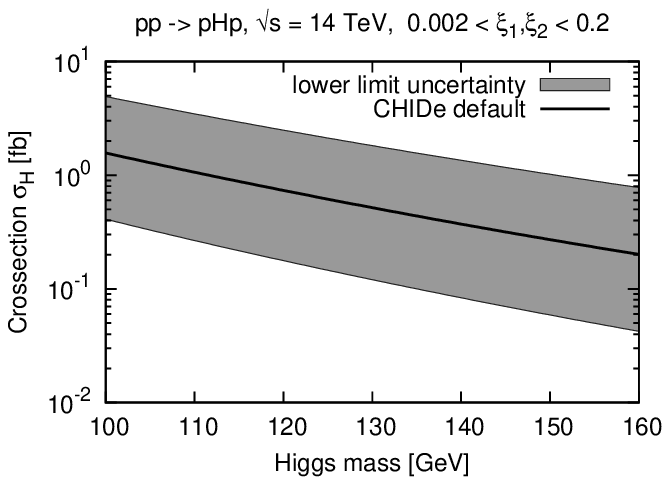}
{Uncertainty due to the lower Sudakov form factor limit for exclusive Higgs at the LHC.}

In this section, we discuss the uncertainties associated with the models of
exclusive diffractive processes. For the analysis we use the CHIDe model,
expecting the results for the KMR model to be qualitatively similar. There are
three main sources of the uncertainties. The first one is the uncertainty on
the gap survival probability which will be measured using the first
LHC data. In this work we 
assume a value of 0.1 at the Tevatron and 0.03 at the LHC~\cite{Khoze:2000wk}. 
An additional source of uncertainty originates 
from the gluon density, which contains the hard and the
soft part. Contrary to the hard part, the soft one is not know precisely and
comes from a phenomenological parametrisation.  The last uncertainty comes from
the limits of the Sudakov integral, which have not yet been fixed by
theoretical calculations (apart from the upper limit for the Higgs case) and
thus are not known precisely.

To check the uncertainty due to the gluon distributions four different
parametrisations of unintegrated skewed gluon densities are used to compute the
exclusive jet and Higgs boson cross sections. These four gluon densities
represent the uncertainty spread due to the present knowledge of unintegrated
parton distribution functions. All of them lead to a fair agreement with the
Tevatron exclusive jet measurement and they lead to an uncertainty of about a factor of
3.5 for jets and 2 for Higgs boson exclusive production at the LHC,  Here we
results for the Higgs case are presented in Fig.~4.

To analyse the uncertainties coming from the Sudakov form factor, both upper
and lower limit of integration were varied by a factor 2.  The study
showed that the effect of changing the upper scale is smaller than for the
lower scale. This is especially true at the LHC energies, where the upper
scale uncertainty can be usually neglected. In Fig.~5 we show the uncertainty
of the lower scale for the Higgs case at the LHC. 

\section{Predictions at the LHC}

To make predictions for exclusive production at the LHC, we need to constrain
the model using the Tevatron data.  The basic idea is to fit the model
parameters to the CDF measurement and use the obtained values at the LHC
energy. We take into consideration both the gluon uncertainty and the dominant,
lower limit of the Sudakov form factor calculation. The principle is simple:
for each gluon density (GLU1 to GLU4), we choose a range of lower limit values
which are compatible with the CDF measurement, taking into account the CDF data
error. The same limit values are used at LHC energies to predict the jet
(Fig.~6) and Higgs (Fig.~7) cross sections. The obtained uncertainty is large,
the factor between the lower and upper edges of the uncertainty is greater than
10 for jets and about 25 for Higgs production.

\sidebyside{h}
{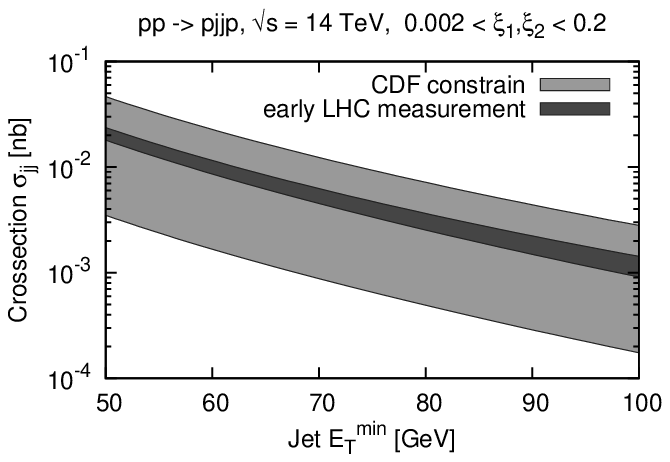}
{Total uncertainty for exclusive jets at the LHC: constraint from the CDF date
and possible LHC data with a low luminosity of 100 pb$^{-1}$.}
{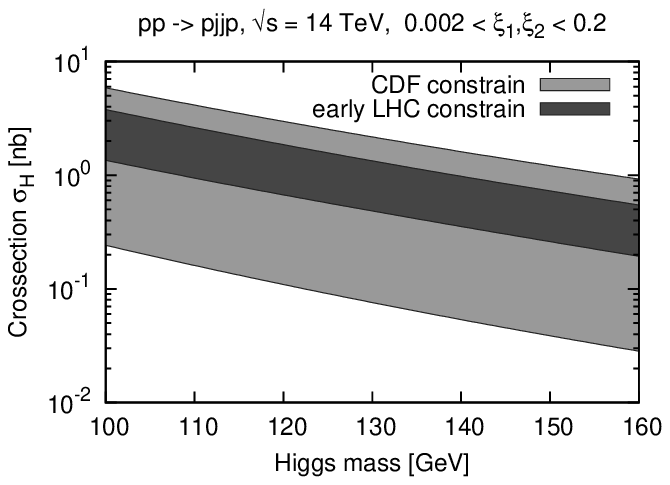}
{Total uncertainty for exclusive Higgs production at the LHC: constraint from
the CDF data and possible LHC data with 100 pb$^{-1}$.}

In order to the previously obtained uncertainty on the Higgs boson cross section, 
we study the possible constraints using early LHC measurement of
exclusive jets -- we assume a possible measurement for 100 pb$^{-1}$. In
addition to the statistical uncertainties, we consider a conservative $3\%$ jet
energy scale uncertainty as the dominant contribution to the systematic error.
A~possible result of such measurement is presented in Fig.~6.  Using the same
prescription as before, we fit the model parameters and obtain the possible
constrained prediction for Higgs (Fig.~7). 

\section{Conclusions} 

Both KMR and CHIDe models describe fairly the CDF measurement of
exclusive jets, but at the LHC energy their predictions differ. This is because
there are several sources of uncertainties of such theoretical description and
in fact the total uncertainty for exclusive production at the LHC is 
large (factor 25). It is possible to constrain
the Higgs boson cross sections within a factor 2 using 
early LHC measurement of exclusive jets.

\end{document}